\documentclass[prx,twocolumn,floatfix,a4paper,superscriptaddress]{revtex4}

\usepackage{bm,color,amsmath,txfonts}
\usepackage{graphicx}
\usepackage{siunitx}
\usepackage{subfigure}
\usepackage{verbatim}
\usepackage{dcolumn}
\usepackage{bm}
\usepackage{epsf}
\usepackage{xcolor}
\usepackage{hyperref}
\usepackage{hhline}
\usepackage{float}
\usepackage{enumerate}
\usepackage{bbm}
\usepackage{lipsum}
\usepackage{mathrsfs}
\usepackage{ulem}

\begin{document}

\title{Cooling mechanical motion with polaritons}

\author{Xuan Zuo}
\affiliation{Zhejiang Key Laboratory of Micro-Nano Quantum Chips and Quantum Control, School of Physics, and State Key Laboratory for Extreme Photonics and Instrumentation, Zhejiang University, Hangzhou 310027, China}
\author{Zi-Xu Lu}
\affiliation{Zhejiang Key Laboratory of Micro-Nano Quantum Chips and Quantum Control, School of Physics, and State Key Laboratory for Extreme Photonics and Instrumentation, Zhejiang University, Hangzhou 310027, China}
\author{Jie Li}\thanks{jieli007@zju.edu.cn}
\affiliation{Zhejiang Key Laboratory of Micro-Nano Quantum Chips and Quantum Control, School of Physics, and State Key Laboratory for Extreme Photonics and Instrumentation, Zhejiang University, Hangzhou 310027, China}

\begin{abstract}
The strong coupling between light and matter gives rise to polaritons. Further coupling polaritons to phonons leads to the formation of hybrid polaromechanical systems. Recent experiments have achieved the strong coupling between polaritons and phonons in two configurations, namely, the magnon-photon-phonon and exciton-photon-phonon systems, which enables the control of mechanical motion via manipulating polaritons.   Here, we present a polaromechanical cooling theory and explicitly show how two polaritons can be used to simultaneously cool two mechanical modes.  The unique advantage of our protocol lies in the fact that the continuous tunability of the polariton frequencies over a wide range allows for the cooling of any two mechanical modes with their frequency difference falling within this range. We further discuss how to extend the theory to cool multiple mechanical modes.  The protocol is designed for cooling mechanical motion in various emerging polaromechanical platforms, such as magnon-, exciton-, and plasmon-polaromechanical systems, which is the first step towards quantum states generation in these hybrid systems. 
\end{abstract}

\maketitle

\section{Introduction}

Polaritons, hybrid quasiparticles arising from the strong coupling between photons and excitations of solid or liquid matter, possess dual characteristics of both light and matter and have emerged as versatile platforms for integrating distinct physical systems~\cite{Hopfield58,Deng,Bellessa,Quirion19,Baranov}, exploring macroscopic quantum phenomena, including Bose-Einstein condensation~\cite{BEC1,BEC2}, Bogoliubov excitations~\cite{Bogoliubov}, quantized vortices~\cite{vortices}, and superfluidity~\cite{fluid}, and designing novel quantum devices, such as spin memory~\cite{Cerna13}, light-emitting diodes~\cite{Tsintzos08,Chakraborty19}, and transistors~\cite{Ballarini13,Zasedatelev19}. 
Further coupling polaritons to more excitations of different nature offers the possibility to build hybrid quantum systems, which combine different physical systems with their individual strengths and complementary functionalities and hold great potential in quantum information science and technologies~\cite{Kurizki15,Naka20}. Recently, two experiments have demonstrated that it is possible to achieve the strong coupling between polaritons and phonons in hybrid exciton-photon-phonon~\cite{Kuznetsov23} and magnon-photon-phonon~\cite{Shen23} systems.  The polaromechanical (PM) strong coupling enables the coherent quantum control of mechanical states via manipulating the polaritons.

Here we show, in such novel PM systems, how polaritons can be used to achieve the cooling of multiple mechanical modes.
 Our PM cooling theory can be readily applied to various PM systems, such as magnon-~\cite{Shen23} and exciton-PM~\cite{Santos23} systems, where the mechanical frequency ranges from megahertz to tens of gigahertz.  In these systems, the PM coupling is realized via, e.g., the deformation potential~\cite{Kuznetsov23}, magnetostrictive~\cite{Zuo24}, and photoelastic~\cite{Santos23} interaction.  A noteworthy advantage of the PM cooling is that, {by exploiting the {\it continuous} tunability of the polariton frequencies over a wide range, it can simultaneously cool multiple mechanical modes whose frequency differences fall within this range.} 
Specifically, by tuning the frequencies of $N$ polaritons to respectively resonate with the scattered mechanical sidebands (associated with $N$ mechanical modes) of the drive field, $N$ anti-Stokes scattering processes are activated leading to the cooling of $N$ mechanical modes all at once.   
The protocol works efficiently in the resolved sideband limit, where the mechanical frequencies and their differences are much larger than the polariton linewidths. 
Impressively, a very small polariton linewidth, ${\sim}10^{-4}$ of the mechanical frequency, has recently been achieved in a magnon-PM system~\cite{Shen23}.
More importantly, it can cool multiple mechanical modes with {\it arbitrary frequencies} once the above condition is fulfilled, benefiting from the great tunability of the polariton eigenfrequencies.
\begin{figure}[b]
	\includegraphics[width=\linewidth]{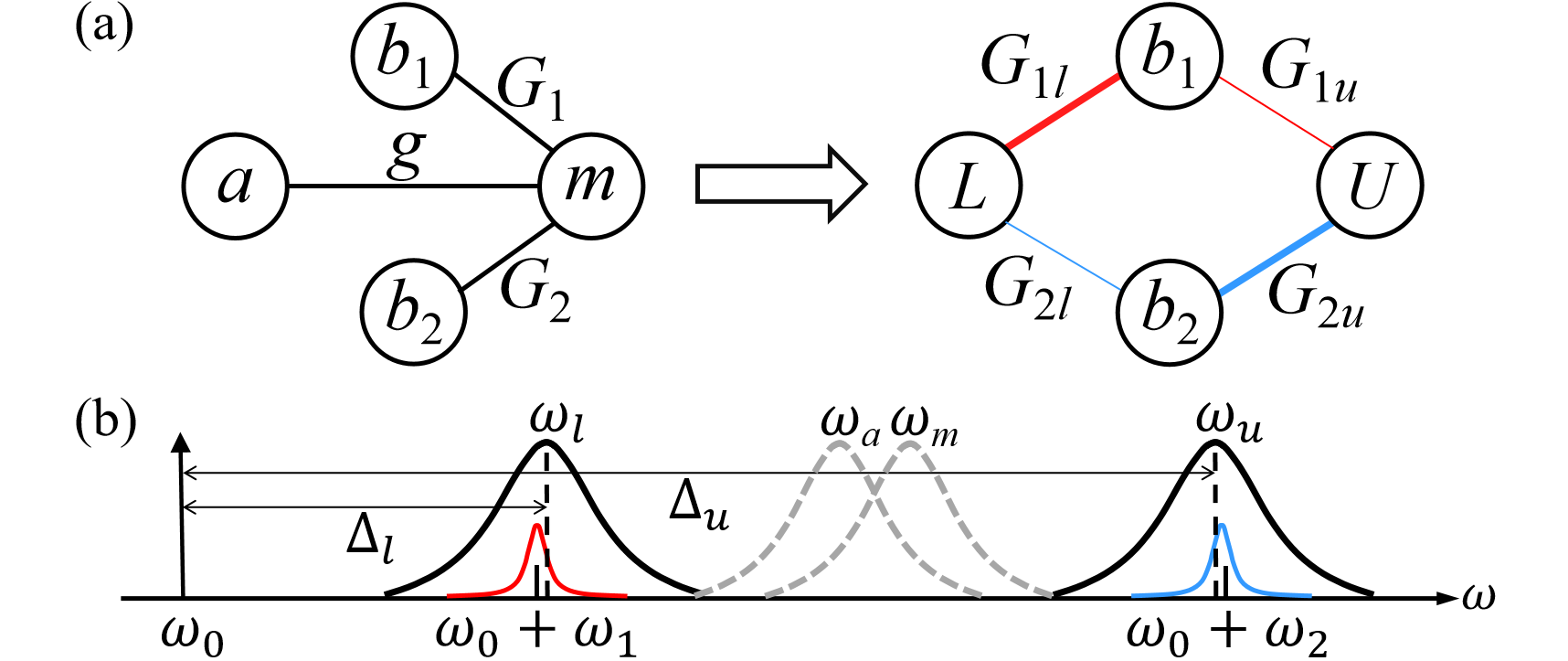}
	\caption{(a) When a microwave cavity mode ($a$) strongly couples to a magnon mode ($m$) forming two magnon polaritons ($L$ and $U$), the cavity magnomechancial system~\cite{Zuo24} becomes a magnon-PM system, where each of the two polaritons couples to two mechanical modes ($b_1$ and $b_2$). (b) Schematic diagram of the protocol for cooling two mechanical modes with two polaritons. Two mechanical motions scatter the driving microwave photons at frequency $\omega_0$ onto two anti-Stokes sidebands at $\omega_0 + \omega_1$ and $\omega_0 + \omega_2$, respectively. When the frequencies of two polaritons are tuned to be resonant with two mechanical sidebands, i.e., $\omega_{l(u)}=\omega_0 + \omega_{1(2)}$, two mechanical modes are simultaneously cooled.}
	\label{fig1}
\end{figure}
A general theory would be a $2N$-mode bosonic system, consisting of $N$ polaritons and $N$ mechanical modes. To illuminate the basic idea of the PM cooling, we specifically start with the simplest case of $N=2$, and take the magnon-PM system as an example. We then discuss how to generalize the results to the case of $N>2$ and to other PM systems.

\section{Polaromechanical system}

A magnon-PM system~\cite{Zuo24,Shen23,Zuo25}, cf.~Fig.~\ref{fig1}(a), consists of a microwave cavity mode ($a$) strongly coupled with a magnon mode ($m$) leading to two magnon polaritons, namely, the upper- (UP) and lower-branch polariton (LP) (denoted as $U$ and $L$ in Fig.~\ref{fig1}(a)), which further couple to two mechanical vibrational modes ($b_1$ and $b_2$) via magnetostriction. Specifically, magnons 
are the collective excitations of a large number of spins in, e.g., an yttrium-iron-garnet (YIG) sphere. They couple to microwave cavity photons via the magnetic dipole interaction~\cite{CM1,CM2,CM3} and to deformation phonons via the magnetostrictive force~\cite{Tang16,Davis21,Jie22,Dong22}. Since the YIG sphere is typically of large size, the frequencies of the vibrational modes are much lower than those of the cavity and magnon modes, promising a dispersive type magnomechanical interaction~\cite{Zuo24}. The Hamiltonian of this four-mode system reads
\begin{align}\label{HHH}
	\begin{split}
		H/\hbar &= \! \omega_{a} a^\dagger a + \omega_{m} m^\dagger m + g \left(a^\dagger m + a m^\dagger \right) {+}\, i\Omega \left(m^\dagger e^{-i\omega_0t} {-} {\rm H.c.} \right) \\& 
		 + \sum_{j=1,2} \bigg( \omega_j b_j^\dagger b_j + G_{0j} m^\dagger m \left(b_j + b_j^\dagger \right) \bigg) ,
	\end{split}
\end{align}
where $a$, $m$, and $b_j$ ($j=1,2$) are the annihilation operators of the cavity, magnon, and two mechanical modes, respectively, satisfying the bosonic commutation relation $[k, k^{\dag}]=1$ ($k = a, m, b_j$), and their resonance frequencies are $\omega_a$, $\omega_m$, and $\omega_j$.  The magnon frequency is tunable over a wide range by varying the external bias magnetic field $H_0$ via $\omega_m = \gamma_0 H_0$, where the gyromagnetic ratio $\gamma_0/2\pi = 28$ GHz/T. The magnon-cavity coupling rate $g$ can be (much) stronger than their dissipation rates, $g > \kappa_a,\, \kappa_m$ (e.g., $g/2\pi =143$ MHz achieved for a $0.8$-mm-diameter YIG sphere~\cite{Tobar}), resulting in two hybridized magnon polaritons~\cite{CM1,CM2,CM3}. Here, $G_{0j}$ denote two bare magnomechanical coupling rates, and the Rabi frequency $\Omega = \frac{\sqrt{5}}{4} \gamma_0 \sqrt{N} B_0$~\cite{Jie18} characterizes the coupling strength between the magnon mode and its drive magnetic field with frequency $\omega_0$ and amplitude $B_0$, where $N = \rho V$ is the number of spins, with $\rho = 4.22 \times 10^{27}$ m$^{-3}$ being the spin density of the YIG  and $V$ the volume of the sphere.

In the strong coupling regime forming two polaritons, the system becomes a four-mode PM system. It is thus convenient to express the Hamiltonian with the polariton operators (Appendix A):
\begin{align}\label{HHHH}
	\begin{split}
		H/\hbar &= \! \omega_{u} U^\dagger U + \omega_{l} L^\dagger L + i\Omega \Big( {\cal M}^\dagger e^{-i\omega_0t} - {\cal M} e^{i\omega_0t} \Big)\\& 
		+ \sum_{j=1,2} \Bigg( \omega_j b_j^\dagger b_j + G_{0j} \big( b_j + b_j^\dagger \big) {\cal M}^\dagger {\cal M} \Bigg) ,
	\end{split}
\end{align}
where $U$ and $L$ are the annihilation operators of the UP and LP, which are the hybridization of the magnon and cavity modes via $U=a \cos\theta + m\sin\theta$ and $L=- a\sin\theta + m\cos\theta$. Here, $ \theta =\frac{1}{2} \arctan \frac{2g}{\Delta_{am}}$ characterizes the mixing ratio of photons and magnons in the polaritons, where $\Delta_{am}\,{=}\,\omega_a\,{-}\,\omega_m$. By altering $\theta$, realized by changing the detuning $\Delta_{am}$ or/and the coupling $g$, the proportions of magnons and cavity photons in the polaritons can be adjusted. 
{A larger coupling $g$ tolerates a larger detuning $\Delta_{am}$ with which the polaritons are present, which implies that the formation of the polaritons does not necessarily require the two original modes to be resonant.}
The eigenfrequencies of the two polaritons are $\omega_{u,l}=\frac{1}{2}\Big(\omega_a + \omega_m \pm \sqrt{\Delta_{am}^2+4g^2} \Big)$. Clearly, the two polariton frequencies can be continuously adjusted by changing $\Delta_{am}$ or $g$. The former can be realized by varying the bias magnetic field $H_0$, and the latter can be easily implemented by changing the position of the YIG sphere inside the microwave cavity. As will be shown later, such tunability of the polariton frequencies (over a wide range) is a core element of our PM cooling, {enabling simultaneous cooling of multiple mechanical modes with frequency differences within this range using a single pump.} 
  For simplicity, we have defined ${\cal M} \equiv U \sin\theta + L \cos\theta$ in Eq.~\eqref{HHHH}. The last coupling term can be expanded as $\sum G_{0j} \big( b_j + b_j^\dagger \big) \left( U^\dagger U \sin^2\theta +  L^\dagger L \cos^2\theta + \sin\theta \cos\theta \left(U^\dagger L + L^\dagger U \right) \right)$, in which the first two terms indicate a dispersive type PM interaction and the third term signifies a phonon-mediated coupling between the two polaritons.

Including dissipations and input noises, the quantum Langevin equations (QLEs) governing the system dynamics are derived, 
which in the frame rotating at the drive frequency are given by (Appendix B)
\begin{align}\label{QLEs}
	\begin{split}
		\dot{U}=&-i\Delta_u U - \kappa_u U - \delta\kappa L  + \Omega \sin\theta + \sqrt[]{2\kappa_u}U^{in}\\& 
		- \sum_{j=1,2} i G_{0j} \left( b_j + b_j^\dagger \right) {\cal M} \sin\theta, \\  
		\dot{L}=&-i\Delta_l L - \kappa_l L - \delta\kappa U + \Omega \cos\theta + \sqrt[]{2\kappa_l}L^{in} \\&
		- \sum_{j=1,2} i G_{0j} \left( b_j + b_j^\dagger \right) {\cal M} \cos\theta , \\ 
		\dot{b}_1=&-i\omega_1 b_1 - \kappa_1 b_1 + \sqrt[]{2\kappa_1} b_1^{in} - i G_{01} {\cal M}^\dagger {\cal M} , \\
		\dot{b}_2=&-i\omega_2 b_2 - \kappa_2 b_2 + \sqrt[]{2\kappa_2} b_2^{in} - i G_{02} {\cal M}^\dagger {\cal M},
	\end{split}
\end{align}
where $\Delta_{u(l)} = \omega_{u(l)}-\omega_0$, $\kappa_u = \kappa_a \cos^2\theta + \kappa_m \sin^2\theta$ and $\kappa_l = \kappa_a \sin^2\theta + \kappa_m \cos^2\theta $ are two polariton dissipation rates, and $\delta \kappa \equiv (\kappa_m-\kappa_a) \sin\theta \cos\theta$ signifies the dissipative coupling between the two polaritons arising from the unbalanced cavity and magnon dissipation rates, which is absent for equal dissipations ($\kappa_m\,\,{=}\,\,\kappa_a$) due to the destructive interference between the dissipation channels. Here, $U^{in} \equiv  (\sqrt[]{2\kappa_a} \cos \theta a^{in} + \sqrt[]{2\kappa_m} \sin \theta m^{in}) /\sqrt[]{2\kappa_u}$ and $L^{in} \equiv (- \sqrt[]{2\kappa_a} \sin \theta a^{in} + \sqrt[]{2\kappa_m} \cos \theta m^{in}) /\sqrt[]{2\kappa_l}$ are the input noise operators of the polaritons, 
$\kappa_{j}$ and $b_{j}^{in}$ are the damping rate and input noise of the mechanical mode $b_{j}$, and $k^{in}(t)$ $(k=a, m, b_1, b_2)$ are zero-mean and characterized by the following correlation functions~\cite{zoller}: $\big\langle k^{in}(t)k^{in\dagger}(t^\prime) \big\rangle=\big[\overline{n}_k(\omega_k)+1 \big]\delta(t-t^\prime)$ and $\big\langle k^{in\dagger}(t)k^{in}(t^\prime) \big\rangle=\overline{n}_k(\omega_k)\delta(t-t^\prime)$, where $\overline{n}_k(\omega_k)=\big[\exp[(\hbar \omega_k/k_BT)]-1\big]^{-1}$ are the mean thermal excitation number of the mode $k$, with the Boltzmann constant $k_B$ and the bath temperature $T$.

The PM coupling stems from the dispersive magnomechanical coupling, which is typically weak for a YIG sphere with the diameter $d>100$ $\mu$m~\cite{Zuo24}. However, an intense microwave drive can significantly enhance the effective magnomechanical (and thus the PM) coupling. The strong drive field results in the two polariton amplitudes $|\langle U \rangle|, |\langle L \rangle| \gg 1$.  This allows us to linearize the system dynamics around the steady-state averages by writing each mode operator as $O=\langle O \rangle + \delta O$ ($O = U, L, b_1, b_2$) and neglecting small second-order fluctuation terms. Substituting the above into Eq.~\eqref{QLEs}, the QLEs are then separated into two sets of equations for classical averages and quantum fluctuations, respectively. 
The steady-state solutions of the averages are given by (Appendix B) 
\begin{align}\label{stav}
	\begin{split}
\langle U \rangle &\approx { -i \Omega \sin\theta}/{{\Delta}_u }, \,\,\,\,\,\,  \langle L \rangle \approx {-i \Omega \cos\theta}/{{\Delta}_l}, \\ 
\langle b_j \rangle &\approx - {G_{0j}} \big| \langle {\cal M} \rangle \big|^2 / {\omega_j}, \,\,\,\,\,  j=1,2.
	\end{split}
\end{align}
The above are obtained under the condition of $\kappa_{u},\kappa_{l}, \delta\kappa \ll \omega_1,\omega_2$, i.e., in the resolved sideband limit, which is well satisfied in the experiments~\cite{Zuo24}.  The linearized QLEs for quantum fluctuations can be written in the following form
\begin{align}\label{uRn}
	\begin{split}
	\dot{u}(t)={\cal R}u(t) + n(t),
	\end{split}
\end{align}
where $u(t)=\big[\delta X_U(t),\delta Y_U(t),\delta X_L(t),\delta Y_L(t),\delta X_{b_1}(t),\delta Y_{b_1}(t),$ $\delta X_{b_2}(t),\delta Y_{b_2}(t) \big]^{\rm T}$ is the vector of quadrature fluctuations, which are defined as $\delta X_O = (\delta O + \delta O^{\dagger})/\!\sqrt{2}$, $\delta Y_O = i (\delta O^{\dagger} - \delta O)/\!\sqrt{2}$; $n(t)=[\sqrt[]{2\kappa_u}X_U^{in},\sqrt[]{2\kappa_U}Y_U^{in}, \sqrt[]{2\kappa_l}X_L^{in},\sqrt[]{2\kappa_l}Y_L^{in},$ $\sqrt[]{2\kappa_1}X_{b_1}^{in},\sqrt[]{2\kappa_1}Y_{b_1}^{in}, \sqrt[]{2\kappa_2}X_{b_2}^{in},\sqrt[]{2\kappa_2}Y_{b_2}^{in}]^{\rm T}$ is the vector of input noises expressed in the quadrature form $X_{O}^{in}$ and $Y_{O}^{in}$; and the drift matrix ${\cal R}$ is given by
\begin{align}
	\cal R=\begin{pmatrix}
		F_u & 0 & A_{1u} & A_{2u} \\
		0 & F_l & A_{1l} & A_{2l} \\
		B_{1u} & B_{1l} & F_1 & 0 \\
		B_{2u} & B_{2l} & 0 & F_2 \\
	\end{pmatrix},
\end{align}
where $F_k$, $F_j$, $A_{jk}$ and $B_{jk}$ ($k = u, l$, $j = 1, 2$) are $2\times2$ matrices: $F_k = \big({-}\kappa_k, {\Delta}_k ; -{\Delta}_k, -\kappa_k \big)$, $F_j = \big({-}\kappa_j, {\omega}_j ; -{\omega}_j, -\kappa_j  \big)$, $A_{ju} = \big({-}G_{jM} \sin\theta, 0 ; 0 , 0  \big)$, $A_{jl} = \big({-}G_{jM} \cos\theta , 0 ; 0 , 0  \big)$, $B_{ju} = \big( 0 , 0 ; 0 , G_{jM} \sin\theta \big)$, and $B_{jl} = \big( 0 , 0 ; 0 , G_{jM} \cos\theta \big)$. Here, $G_{jM}= i 2 G_{0j} \langle {\cal M} \rangle$ is the effective magnomechanical coupling strength associated with the mechanical mode $b_j$. {The analysis of steady-state effective phonon numbers requires that the real parts of the eigenvalues of the drift matrix $\mathcal{R}$ be negative. Thus, the stability conditions are explicitly verified in the following section.}

\section{Polaromechanical cooling of two mechanical modes}

The phonons in each mode scatter the microwave driving photons onto two mechanical sidebands, i.e., the Stokes and anti-Stokes photons. 
The key operation to simultaneously cool two mechanical modes is to adjust the polariton frequencies to be resonant with the two anti-Stokes sidebands, i.e., $\omega_{l(u)} = \omega_0 +  \omega_{1(2)}$ (Fig.~\ref{fig1}(b)), while the two Stokes sidebands at $\omega_0 -  \omega_{1(2)}$ are deeply suppressed in the limit $\kappa_{u}, \kappa_{l} \ll \omega_{1},\omega_{2}$. This leads to a net cooling rate for each mechanical mode, and ground-state cooling can be achieved when it is significantly larger than $\kappa_{1(2)}$.   
Applying the sideband cooling theory~\cite{TJK07,FM07,Genes08} to our PM system, for the mechanical mode $b_1$ and in the weak coupling regime $G_{1M} \ll \kappa_{k}$, we obtain the following two sets of Stokes and anti-Stokes scattering rates $A_{k\pm}^{(1)}$ associated with the two polaritons ($k=u,l$):
\begin{align}
	\begin{split}\label{rate}
	A_{u\pm}^{(1)} = \frac{\kappa_u \, G_{1M}^2 \sin^2\theta}{4 \big[\kappa_u^2 + (\Delta_u \pm \omega_1)^2 \big]} , \quad
	A_{l\pm}^{(1)} = \frac{\kappa_l \, G_{1M}^2 \cos^2\theta}{4 \big[\kappa_l^2 + (\Delta_l \pm \omega_1)^2 \big]} ,
	\end{split}
\end{align}
where $A_{u+}^{(1)}$ and $A_{l+}^{(1)}$ \Big($A_{u-}^{(1)}$ and $A_{l-}^{(1)}$\Big) denote the Stokes (anti-Stokes) scattering rates related to the UP and LP, respectively, and $\sin\theta$ and $\cos\theta$ determine the weights of the magnomechanical coupling strength $G_{1M}$ assigned to the two PM couplings. Since the anti-Stokes (Stokes) scattering cools (amplifies) the mechanical motion, the net cooling rate equals to $A_{-}^{(1)}-A_{+}^{(1)}$. Because both the polaritons contribute to the cooling of the mechanical mode $b_1$, we thus have 
\begin{align}
	\begin{split}
	\Delta \kappa_{1u} = A_{u-}^{(1)}-A_{u+}^{(1)}, \quad \,\,\,
	\Delta \kappa_{1l} = A_{l-}^{(1)}-A_{l+}^{(1)},
	\end{split}
\end{align}
where $\Delta \kappa_{1u}$ ($\Delta \kappa_{1l}$) $>0$ is the net cooling rate due to the dispersive interaction with the UP (LP). The cooling effect is manifested by the polaromechanically induced broadening of the mechanical linewidth, i.e., the effective damping rate 
\begin{align}
\kappa_{\rm 1,eff} = \kappa_1 + \Delta \kappa_{1u} + \Delta \kappa_{1l}, 
\end{align}
which contains the contribution from the two polaritons.

In our protocol, two mechanical modes with appreciably different frequencies $|\omega_1 -\omega_2| \gg \kappa_{u}, \kappa_{l} $ are chosen, and the polariton-drive detunings are set to $\Delta_l = \omega_1$ and $\Delta_u = \omega_2$. In this situation, the LP is resonant with the anti-Stokes sideband of the mechanical mode $b_1$, while the UP is far detuned from it. Thus, the cooling of the mechanical mode is mainly contributed by the LP. This can be verified using Eq.~\eqref{rate} with $\Delta_u = \omega_2$ and $\Delta_l = \omega_1$, and one finds that $A_{l-}^{(1)} \gg A_{u-}^{(1)}, A_{l+}^{(1)}, A_{u+}^{(1)}$ when $\kappa_{u}, \kappa_{l} \ll \omega_{1},\omega_{2}, |\omega_1 -\omega_2|$. Neglecting the small contribution from the UP, the steady-state effective phonon number is given by
\begin{align}\label{n1eff}
	\begin{split}
	n_{\rm 1,eff} \approx \frac{\kappa_1 \overline{n}_{b_1} + A_{l+}^{(1)}}{\kappa_1 + \Delta \kappa_{1l}},
	\end{split}
\end{align}
which coincides with the result of sideband cooling of a single mechanical oscillator~\cite{TJK07,FM07,Genes08}. 

A similar analysis for the second mechanical mode can be done. The UP resonant with the anti-Stokes sideband results in the corresponding anti-Stokes scattering rate as the primary contribution, i.e., $A_{u-}^{(2)} \gg A_{l-}^{(2)}, A_{l+}^{(2)}, A_{u+}^{(2)}$. The definitions of the above scattering rates are the same as in Eq.~\eqref{rate}, but replace the subscript 1 with 2.  Omitting the negligible impact of the LP, the effective mechanical damping rate $\kappa_{\rm 2,eff} \approx \kappa_2 + \Delta \kappa_{2u}$, and the steady-state effective phonon number reads
\begin{align}\label{n2eff}
	\begin{split}
	n_{\rm 2,eff} \approx \frac{\kappa_2 \overline{n}_{b_2} + A_{u+}^{(2)}}{\kappa_2 + \Delta \kappa_{2u}}.
	\end{split}
\end{align}

\begin{figure}[t]
	\includegraphics[width=0.93\linewidth]{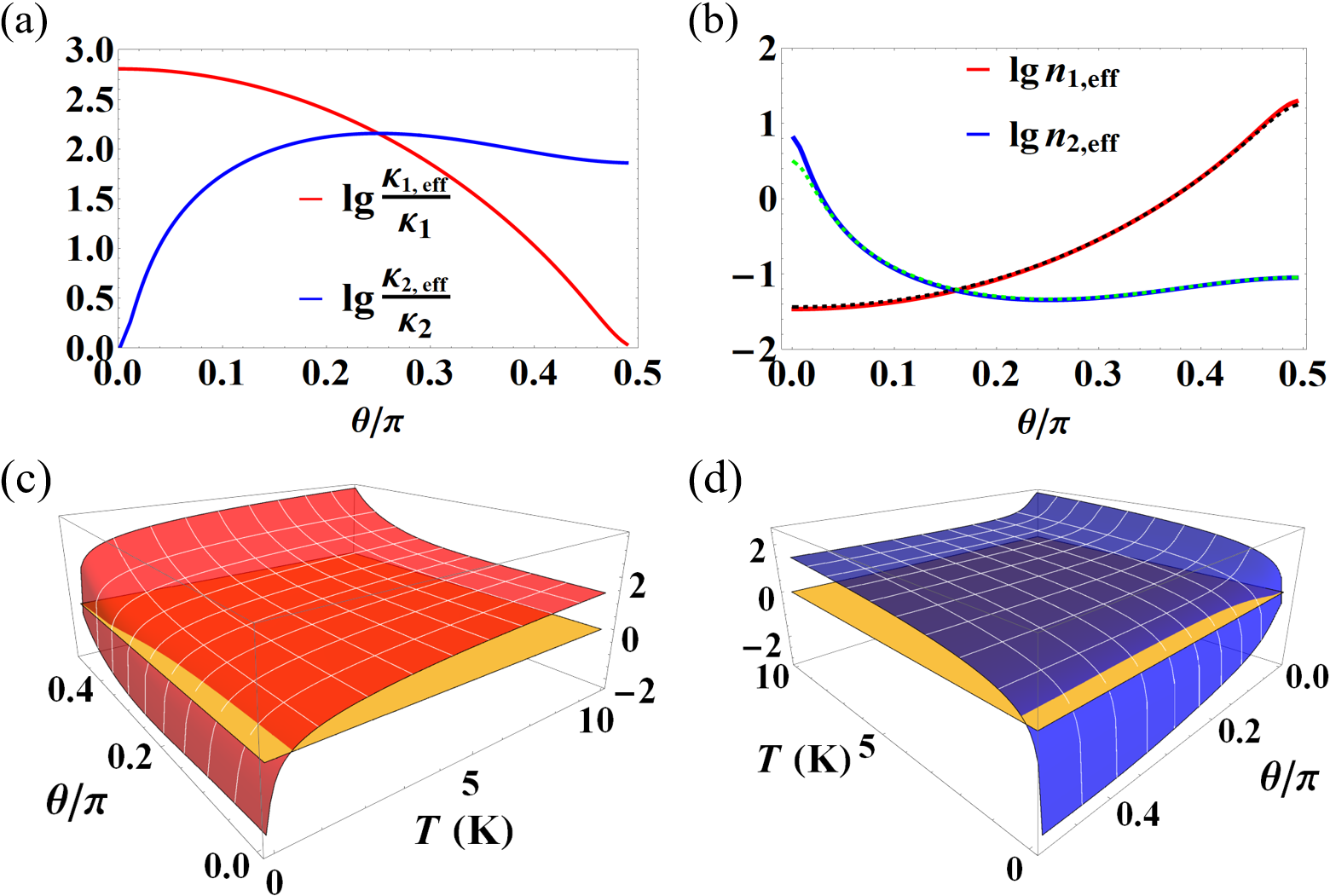}
	\caption{(a) Effective mechanical damping rates $\kappa_{\rm j,eff}$ and (b) phonon numbers $n_{\rm j,eff}$ versus $\theta$. (c) $n_{\rm 1,eff}$ and (d) $n_{\rm 2,eff}$ versus $\theta$ and $T$. They are plotted on a logarithmic scale. In (b), the solid curves are the analytical results of Eqs.~\eqref{n1eff}-\eqref{n2eff} and the dashed curves are the numerical results.  The yellow horizontal planes in (c)-(d) indicate $n_{\rm j,eff}=1$. We take $\Delta_u = \omega_2$ and $\Delta_l = \omega_1$. See text for the other parameters.} 
\label{fig2}
\end{figure}

From the above analysis, one sees that when the resonant conditions $\Delta_u = \omega_2$ and $\Delta_l = \omega_1$, and $\kappa_{u,l} \ll \omega_{1,2}, |\omega_1 -\omega_2|$ are satisfied, the cooling of two mechanical modes can be treated as two independent sideband cooling processes, i.e., each polariton is responsible for cooling one mechanical motion. This reveals the high efficiency of our protocol, i.e., multiple mechanical modes can be simultaneously cooled by exploiting multiple {\it tunable} polaritons. 
Besides, there is competition between the above two cooling processes, which originates from the competition relation in distributing magnons into the two polaritons (with weights of $\sin^2\theta$ and $\cos^2\theta$), which further leads to the competition between the two PM coupling rates (cf., Eq.~\eqref{rate}) and thus between the two anti-Stokes scattering rates $A_{l-}^{(1)}$ and $A_{u-}^{(2)}$.
This is seen in Fig.~\ref{fig2}(a)-\ref{fig2}(b), where a small $\theta \to 0$ ($\theta \to \frac{\pi}{2}$) corresponds to a large proportion of magnons and thus magnomechanical coupling strength in the LP (UP), and since the LP (UP) is responsible for cooling the lower (higher)-frequency mechanical mode, the mode $b_1$ ($b_2$) gets efficiently cooled, indicated by the significantly increased damping rate $\kappa_{\rm j,eff}/\kappa_j \gg 1$ (Fig.~\ref{fig2}(a)) and reduced phonon number $n_{\rm j,eff} /\overline{n}_{b_j} \ll 1$ (Fig.~\ref{fig2}(b)). 
Nevertheless, for moderate values of $\theta$, both the mechanical modes can be cooled into their quantum ground state with $n_{\rm eff} \ll1$. 
In Fig.~\ref{fig2}(b), the analytical results of Eqs.~\eqref{n1eff}-\eqref{n2eff} (solid curves) agree well with the numerical results (dashed curves) obtained by solving the Lyapunov equation associated with Eq.~\eqref{uRn} for the steady-state energies of two mechanical modes (Appendix C), implying two independent cooling processes in the limit of resolved sidebands and large mechanical frequency difference.   Note that the red and blue curves in Figs.~\ref{fig2}(a)-\ref{fig2}(b) are asymmetrical about $\theta=\frac{\pi}{4}$, which is fundamentally due to different mechanical frequencies.  
We have employed experimentally feasible parameters~\cite{Zuo24,Shen23}: $\omega_a \,{=}\, 2\pi \times 10$~GHz, $\omega_2 \,{=}\, 3 \omega_1 \,{=}\, 2\pi \times 30$~MHz, $\kappa_{a} = \kappa_{m} =2\pi \times 1$~MHz (giving $\delta \kappa =0$), $\kappa_{j} = 2\pi \times 10^2$~Hz, $G_{0j} = 2\pi \times 0.2$~Hz $(j=1,2)$, $T = 10$~mK, and the amplitude of the drive field $B_0 = 2.7 \times 10^{-5}$~T, corresponding to the drive power $P = 4.3$~mW for a 250-$\mu$m-diameter YIG sphere. 
The protocol works efficiently for higher bath temperatures, as shown in Figs.~\ref{fig2}(c)-\ref{fig2}(d), where both the modes can be cooled to $n_{\rm eff}<1$ for $T$ being up to $\sim0.5$ K. We use a higher drive power $P = 69$ mW {in getting Figs.~\ref{fig2}(c)-\ref{fig2}(d)}, which yields stronger PM couplings and thus larger cooling rates. In this case, the weak coupling condition for deriving Eqs.~\eqref{n1eff}-\eqref{n2eff} is broken, so Figs.~\ref{fig2}(c)-\ref{fig2}(d) are plotted numerically.

{It is worth mentioning that in getting Fig.~\ref{fig2}, $\theta$ is altered to reveal the mechanism of the PM cooling, which is achieved by varying the cavity-magnon detuning $\Delta_{am}\,{=}\,\omega_a\,{-}\,\omega_m$ from positive to negative (specifically, by continuously increasing the magnon frequency $\omega_m$ since the cavity frequency $\omega_a$ is fixed). Meanwhile, the magnon-cavity coupling rate $g$ should also be adjusted to satisfy the condition $\omega_u - \omega_l \equiv \sqrt{\Delta_{am}^2 + 4g^2} = |\omega_2 - \omega_1|$, which ensures that the frequency difference between the two polaritons matches that of the two mechanical modes (Fig.~\ref{fig1}(b)). Finally, the drive frequency $\omega_0$ is set as $\omega_0 = \frac{\omega_a + \omega_m}{2} - \frac{\omega_1 + \omega_2}{2}$ to satisfy the optimal cooling condition, i.e., the two mechanical sidebands being resonant with the two polariton modes, $\Delta_{u(l)} = \omega_{2(1)}$.}

\begin{figure}[t]
	\includegraphics[width=0.93\linewidth]{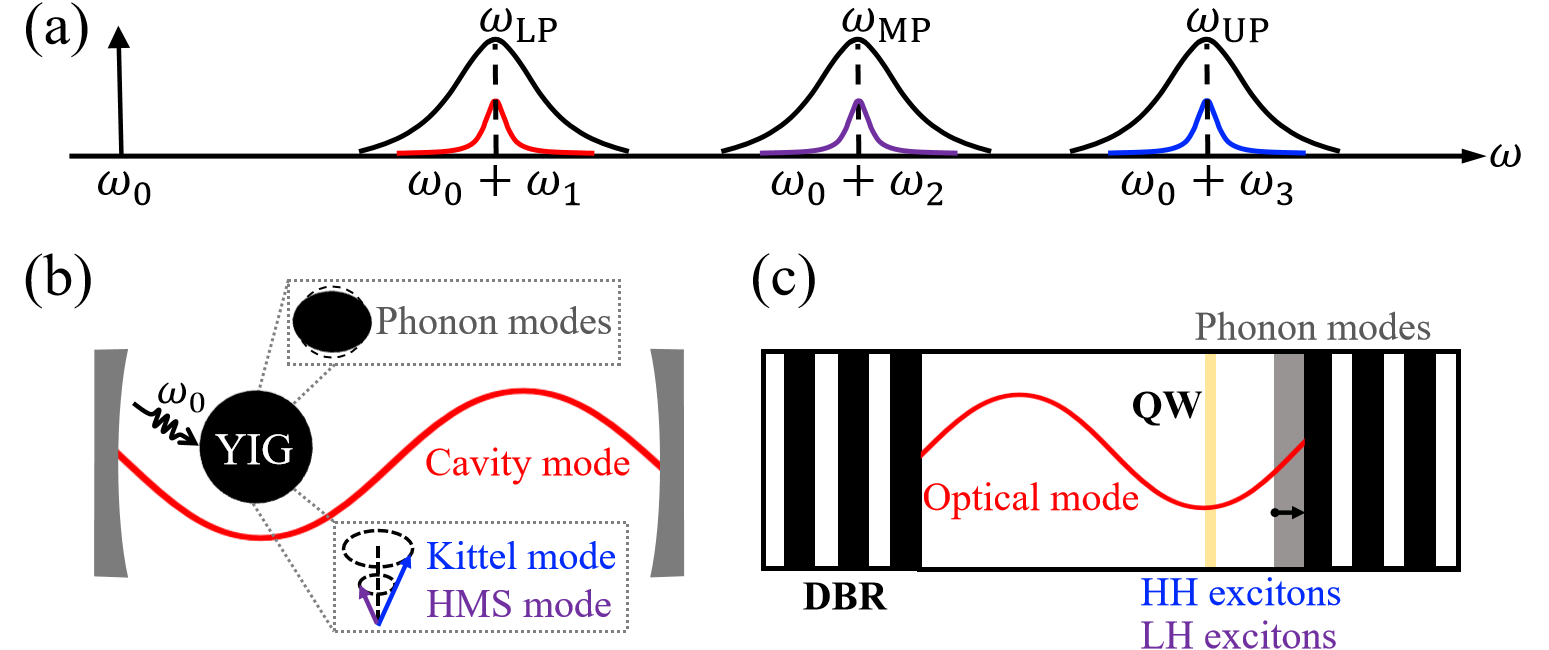}
	\caption{(a) Schematic diagram of the protocol for cooling three mechanical modes with three polaritons (LP, MP and UP). The proposed (b) magnon- or (c) exciton-PM system to realize the protocol.}
	\label{fig3}
\end{figure}

\section{Discussion and Conclusion}

For the cooling of multiple mechanical modes, e.g., three mechanical modes, three polariton modes should be employed, which resonantly enhance the corresponding three anti-Stokes scatterings (Fig.~\ref{fig3}(a)). Three magnon polaritons can be achieved by strongly coupling the microwave cavity to two magnonic modes of a YIG sphere, e.g., the Kittel and a high-order magnetostatic (HMS) mode~\cite{Wu22} (Fig.~\ref{fig3}(b)). More polaritons can be realized by coupling to more HMS modes, which can be used to cool a series of vibrational modes. The protocol can also be applied to exciton-PM systems~\cite{Santos23,Kyriienko14,Bloch22,Sesin23} (Fig.~\ref{fig3}(c)), where an optical cavity mode strongly couples to two exciton modes, e.g., heavy- (HH) and light-hole (LH) excitons, in a semiconductor microcavity, leading to three exciton polaritons, i.e., lower (LP), middle (MP), and upper (UP) polaritons~\cite{Sesin23}. They can be exploited to cool three highly monochromatic phonon modes in the GHz range~\cite{Fainstein13} using the protocol depicted in Fig.~\ref{fig3}(a).   In this system, the polariton frequency is also tunable as the semiconductor microcavity features a thickness taper that enables the variation of the cavity frequency~\cite{Weisbuch}, and the exciton-photon coupling strength can be adjusted by changing the position of the quantum well (QW) within the cavity. Our protocol can potentially be applied to other developing PM systems, e.g., the plasmon-PM system realized by coupling plasmons to an optomechancial microcavity.   
It is general in the sense that the PM coupling can be achieved in various systems via, e.g., the magnon-~\cite{Zuo24} and exciton-phonon~\cite{Li24} interactions.

In conclusion, we have shown how to use polaritons to cool mechanical modes, potentially to their quantum ground state. By exploiting the flexible tunability of the polariton frequencies, our protocol holds potential for cooling multiple mechanical modes in a variety of PM systems.  {Ground-state cooling of mechanical motion is a prerequisite for preparing quantum states in PM systems. The work provides guidance for the experimental realization of mechanical cooling with polaritons, which may stimulate the experiments towards quantum states generation in the newly-developed PM platforms. }


\section*{ACKNOWLEDGMENTS}
We thank D. Vitali, F. Massel, and S. Zippilli for helpful discussions and feedback on the manuscript.  This work was supported by National Key Research and Development Program of China (2022YFA1405200, 2024YFA1408900), National Natural Science Foundation of China (12474365, 92265202), and Zhejiang Provincial Natural Science Foundation of China (LR25A050001).

\section*{APPENDIX A: POLAROMECHANICAL HAMILTONIAN}\label{appA}

\setcounter{figure}{0}
\setcounter{equation}{0}
\setcounter{table}{0}
\renewcommand\theequation{A\arabic{equation}}
\renewcommand\thefigure{A\arabic{figure}}
\renewcommand\thetable{A\arabic{table}}

The Hamiltonian of the cavity magnomechanical system, involving a microwave cavity mode, a magnon mode and  two mechanical modes, reads~\cite{Zuo24}
\begin{align}\label{HHHA}
	\begin{split}
		H/\hbar &= \! \omega_{a} a^\dagger a + \omega_{m} m^\dagger m + g \left(a^\dagger m + a m^\dagger \right)\\
		&+ i\Omega \left(m^\dagger e^{-i\omega_0t} {-}\, m e^{i\omega_0t} \right) {+} \sum_{j=1,2} \bigg( \omega_j b_j^\dagger b_j {+}\, G_{0j} m^\dagger m \left(b_j + b_j^\dagger \right) \bigg) ,
	\end{split}
\end{align}
where the Hamiltonian of the cavity-magnon subsystem can be diagonalized as follows
\begin{align}
	\begin{split}
		H_{a-m}/\hbar &= \! \omega_{a} a^\dagger a + \omega_{m} m^\dagger m + g \left(a^\dagger m+am^\dagger \right)\\
	     		 &= 	
	\begin{pmatrix}a^\dagger & m^\dagger\end{pmatrix}
	\begin{pmatrix}\omega_a & g \\ g & \omega_m\end{pmatrix}
	\begin{pmatrix}a\\	m\end{pmatrix}\\
	     		 &= 	
	\begin{pmatrix}a^\dagger & m^\dagger\end{pmatrix}F^{-1}
	\begin{pmatrix}\omega_u & 0\\0 & \omega_l\end{pmatrix}F
	\begin{pmatrix}	a\\m\end{pmatrix}\\
	     		 &= 	
	\begin{pmatrix}U^\dagger & L^\dagger\end{pmatrix}
	\begin{pmatrix}\omega_u & 0\\0 & \omega_l\end{pmatrix}
	\begin{pmatrix}	U\\L\end{pmatrix}\\
					 &= \! \omega_{u} U^\dagger U + \omega_{l} L^\dagger L  .
	\end{split}
\end{align}
This corresponds to the formation of two cavity-magnon polaritons in the strong coupling regime, i.e., the upper- and the lower-branch polariton with the corresponding annihilation operators $U$ and $L$. Their eigenfrequencies are given by
\begin{align}\label{eigenfreq}
	\begin{split}
	\omega_u & = \frac{\omega_a+\omega_m+\sqrt{(\omega_a-\omega_m)^2+4g^2}}{2},\\
	\omega_l & = \frac{\omega_a+\omega_m-\sqrt{(\omega_a-\omega_m)^2+4g^2}}{2}.
	\end{split}
\end{align}
The diagonalization transformation matrix
\begin{align}
	\begin{split}
		F = 	\begin{pmatrix} \cos\theta & \sin\theta\\ -\sin\theta & \cos\theta \end{pmatrix}
	\end{split}
\end{align}
indicates that the two polaritons are formed by the hybridization of cavity photons and magnons via
\begin{align}\label{transformation}
	\begin{split} 
		U &= a \cos\theta + m\sin\theta, \\ 
		L &= - a\sin\theta + m\cos\theta,
	\end{split}
\end{align} 
where $\theta = \frac{1}{2} \arctan \frac{2g}{\omega_a - \omega_m}$ ($\theta \in [0,\frac{\pi}{2}]$) characterizes the mixing ratio of the both. It can be verified that the creation and annihilation operators of the two polaritons still satisfy the bosonic commutation relation $[O,O^\dagger]=1$ ($O=U, L$).

In the strong coupling regime, the system becomes a four-mode polaromechanical system. We thus express the Hamiltonian~\eqref{HHHA} in terms of two polariton operators, i.e.,
\begin{align}
	\begin{split}
		H/\hbar &= \! \omega_{u} U^\dagger U + \omega_{l} L^\dagger L  \\&
		+ i\Omega \Big[ \big( U^\dagger \sin\theta + L^\dagger \cos\theta \big) e^{-i\omega_0t} - \big( U \sin\theta + L \cos\theta \big) e^{i\omega_0t} \Big]\\& 
		+ \sum_{j=1,2} \Bigg\{ \omega_j b_j^\dagger b_j + G_{0j} \big( b_j + b_j^\dagger \big) \\& 
		\times \Big[U^\dagger U \sin^2\theta + L^\dagger L \cos^2\theta  +  \frac{1}{2} \big(U^\dagger L  + L^\dagger U \big) \sin2\theta \Big] \Bigg\},
	\end{split}
\end{align}
which clearly shows that each of the two polaritons couples to both the mechanical modes via a polaromechanical interaction, which is a dispersive type (radiation pressure-like)
interaction.  By defining ${\cal M} \equiv U \sin\theta + L \cos\theta$, the above Hamiltonian reduces to a more compact form
\begin{align}\label{HHHP}
	\begin{split}
		H/\hbar &= \! \omega_{u} U^\dagger U + \omega_{l} L^\dagger L + i\Omega \Big( {\cal M}^\dagger e^{-i\omega_0t} - {\cal M} e^{i\omega_0t} \Big) \\& 
		+ \sum_{j=1,2} \Bigg( \omega_j b_j^\dagger b_j + G_{0j} \big( b_j + b_j^\dagger \big) {\cal M}^\dagger {\cal M} \Bigg) .
	\end{split}
\end{align}

\section*{APPENDIX B: LINEARIZED QUANTUM LANGEVIN EQUATIONS}
\label{appB}

\setcounter{figure}{0}
\setcounter{equation}{0}
\setcounter{table}{0}
\renewcommand\theequation{B\arabic{equation}}
\renewcommand\thefigure{B\arabic{figure}}
\renewcommand\thetable{B\arabic{table}}

The quantum Langevin equations (QLEs) for the polaromechanical system can be derived by making the transformation from 
the QLEs of the original modes. The cavity magnomechanical Hamiltonian~\eqref{HHHA} gives rise to the following QLEs in the frame rotating at the drive frequency $\omega_0$:
\begin{align}~\label{QLEsO}
	\begin{split}
		\dot{a}=&-(i\Delta_a + \kappa_a) a - i g m + \sqrt[]{2\kappa_a}a^{in},  \\
		\dot{m}=&-(i\Delta_m + \kappa_m) m - i g a - \sum_{j=1,2} i G_{0j} m \left(b_j + b_j^\dagger \right) + \Omega + \sqrt[]{2\kappa_m}m^{in}, \\
		\dot{b}_1=&-i\omega_1 b_1 - \kappa_1 b_1 + \sqrt[]{2\kappa_1} b_1^{in} - i G_{01} m^\dagger m , \\
		\dot{b}_2=&-i\omega_2 b_2 - \kappa_2 b_2 + \sqrt[]{2\kappa_2} b_2^{in} - i G_{02} m^\dagger m,
	\end{split}
\end{align}
where $\Delta_{a} = \omega_{a}-\omega_0$ and $\Delta_{m} = \omega_{m}-\omega_0$. Applying the transformation Eq.~\eqref{transformation} to the above QLEs, we obtain
\begin{align}
	\begin{split}
		\dot{U}=&-i\Delta_u U - \kappa_u U - \delta\kappa L  + \Omega \sin\theta + \sqrt[]{2\kappa_u}U^{in}\\& 
		- \sum_{j=1,2} i G_{0j} \left( b_j + b_j^\dagger \right) {\cal M} \sin\theta, \\  
		\dot{L}=&-i\Delta_l L - \kappa_l L - \delta\kappa U + \Omega \cos\theta + \sqrt[]{2\kappa_l}L^{in} \\&
		- \sum_{j=1,2} i G_{0j} \left( b_j + b_j^\dagger \right) {\cal M} \cos\theta , \\ 
		\dot{b}_1=&-i\omega_1 b_1 - \kappa_1 b_1 + \sqrt[]{2\kappa_1} b_1^{in} - i G_{01} {\cal M}^\dagger {\cal M} , \\
		\dot{b}_2=&-i\omega_2 b_2 - \kappa_2 b_2 + \sqrt[]{2\kappa_2} b_2^{in} - i G_{02} {\cal M}^\dagger {\cal M},
	\end{split}
\end{align}
where $\Delta_{u} = \omega_{u}-\omega_0$ and $\Delta_{l} = \omega_{l}-\omega_0$ are the polariton-drive detunings, and 
$\kappa_u = \kappa_a \cos^2\theta + \kappa_m \sin^2\theta$ and $\kappa_l = \kappa_a \sin^2\theta + \kappa_m \cos^2\theta $ are two polariton dissipation rates.  $\delta \kappa \equiv (\kappa_m-\kappa_a) \sin\theta \cos\theta$ denotes the dissipative coupling between the two polaritons due to the unbalanced cavity and magnon decay rates $\kappa_m \neq \kappa_a$, which is negligibly small when the two decay rates are close.  
$U^{in} \equiv  (\sqrt[]{2\kappa_a} \cos \theta a^{in} + \sqrt[]{2\kappa_m} \sin \theta m^{in}) /\sqrt[]{2\kappa_u}$ and $L^{in} \equiv (- \sqrt[]{2\kappa_a} \sin \theta a^{in} + \sqrt[]{2\kappa_m} \cos \theta m^{in}) /\sqrt[]{2\kappa_l}$ represent the input noises for the two polaritons, which are defined to be consistent with the standard form as in Eq.~\eqref{QLEsO}.

Under a strong drive field, the amplitudes of two polaritons in the steady state $|\langle U \rangle|, |\langle L \rangle| \gg 1$. This allows us to linearize the dynamics around the steady-state averages by writing each mode operator as $O=\langle O \rangle + \delta O$ ($O = U, L, b_1, b_2$) and neglecting small second-order fluctuation terms. The linearized QLEs for the quantum fluctuations are given by
\begin{align}\label{qleULB}
	\begin{split}
		\dot{\delta U}{=}&\,{-} \big( i\tilde{\Delta}_u + \kappa_u \big) \delta U \,{-}\, \big(i {G_{ul}} + \delta\kappa \big) \delta L \\ &{-}\sum_{j=1,2} G_{jM} {\sin \theta} \frac{\delta b_j + \delta b_j^\dagger}{2} \,{+}\, \sqrt[]{2\kappa_u}A_u^{in},  \\
		\dot{\delta L}{=}&\,{-} \big( i\tilde{\Delta}_l + \kappa_l \big) \delta L \,{-}\, \big(i {G_{ul}} + \delta\kappa \big) \delta U \\&{-}\sum_{j=1,2} G_{jM} {\cos \theta} \frac{\delta b_j + \delta b_j^\dagger}{2} \,{+}\, \sqrt[]{2\kappa_l}L^{in},  \\
		\dot{\delta b_1}\,{=}& \,{-} \big( i \omega_1 + \kappa_1 \big) \delta b_1 \,{+}\, \sqrt[]{2\kappa_1}b_1^{in}  \\&{-}\, {\left(\frac{G_{1M} {\sin \theta}}{2} \delta U^\dagger + \frac{G_{1M} {\cos \theta}}{2} \delta L^\dagger - \rm{H.\,c.} \right)},	\\
		\dot{\delta b_2}\,{=}& \,{-} \big( i \omega_2 + \kappa_2 \big) \delta b_2 \,{+}\, \sqrt[]{2\kappa_2}b_2^{in} \\&{-} {\left(\frac{G_{2M} {\sin \theta}}{2} \delta U^\dagger + \frac{G_{2M} {\cos \theta}}{2} \delta L^\dagger - \rm{H.\,c.} \right)},	
	\end{split}
\end{align}
where $\tilde{\Delta}_u  = \Delta_u + \sum_{j=1,2} 2 G_{0j} {\rm Re} \langle b_j \rangle \sin^2\theta$ and $\tilde{\Delta}_l  = \Delta_l + \sum_{j=1,2} 2 G_{0j} {\rm Re} \langle b_j \rangle \cos^2\theta$ are the effective detunings including the frequency shift induced by the polaromechanical interaction. $G_{jM}= i 2 G_{0j} \langle {\cal M} \rangle$ is the effective magnomechanical coupling strength associated with the mechanical mode $b_j$, and $G_{ul} = \sum_{j=1,2} G_{0j} {\rm Re} \langle b_j \rangle \sin 2 \theta$ denotes the phonon-mediated coupling between the two polaritons. Since the bare magnomechanical coupling rate $G_{0j}$ is typically small~\cite{Zuo24} and $G_{ul} \propto G_{0j}^2$ (given $\langle b_j \rangle$ $\propto$ $G_{0j}$ as shown below in Eq.~\eqref{bbbb}), hereafter we neglect this phonon-mediated indirect coupling. For simplicity, we also neglect the dissipative coupling due to $\delta \kappa$ as the two decay rates $\kappa_a$ and $\kappa_m$ can be designed very close in the cavity magnonics experiments. In our protocol, the two polariton-drive detunings are set to be equal to the mechanical frequencies, i.e., $\tilde{\Delta}_u =\omega_2$ and $\tilde{\Delta}_l =\omega_1$, which are much larger than the polaromechanically induced frequency shift, so in the following we can safely assume $\tilde{\Delta}_u \approx \Delta_u$ and $\tilde{\Delta}_l \approx \Delta_l$. 

The steady-state averages of the two polaritons are given by
\begin{align}
	\begin{split}
		\langle U \rangle &= \frac{\delta\kappa \Omega \cos \theta - i \Omega \sin \theta({\Delta}_l - i \kappa_l)}{({\Delta}_l - i \kappa_l)({\Delta}_u - i \kappa_u) + \delta\kappa^2},\\ 
		\langle L \rangle &= \frac{\delta\kappa \Omega \sin \theta - i \Omega \cos \theta({\Delta}_u - i \kappa_u)}{({\Delta}_l - i \kappa_l)({\Delta}_u - i \kappa_u) + \delta\kappa^2}.
	\end{split}
\end{align}
Under the condition of $\kappa_u, \kappa_l, \delta \kappa \ll \omega_1, \omega_2$, i.e., in the resolved sideband limit, the above averages can be simplified as 
\begin{align}
	\begin{split}
		\langle U \rangle &\approx \frac{ -i \Omega \sin\theta}{{\Delta}_u } ,\\ 
		\langle L \rangle &\approx \frac{-i \Omega \cos\theta}{{\Delta}_l}.
			\end{split}
\end{align}
The steady-state averages of two mechanical modes can be approximated in the same manner, i.e.,
\begin{align}\label{bbbb}
	\begin{split}
		\langle b_1 \rangle &= - \frac{{G_{01}} \big| \langle {\cal M} \rangle \big|^2} {\omega_1 - i \kappa_1} \approx - \frac{{G_{01}} \big| \langle {\cal M} \rangle \big|^2} {\omega_1},\\
		\langle b_2 \rangle &= - \frac{{G_{02}} \big| \langle {\cal M} \rangle \big|^2} {\omega_2 - i \kappa_2} \approx - \frac{{G_{02}} \big| \langle {\cal M} \rangle \big|^2} {\omega_2}.
	\end{split}
\end{align}

\section*{APPENDIX C: STEADY-STATE EFFECTIVE PHONON NUMBERS}
\label{appC}

\setcounter{figure}{0}
\setcounter{equation}{0}
\setcounter{table}{0}
\renewcommand\theequation{C\arabic{equation}}
\renewcommand\thefigure{C\arabic{figure}}
\renewcommand\thetable{C\arabic{table}}

The steady-state effective phonon numbers can be extracted from the steady-state energies of the two mechanical modes, which we calculate below using the covariance-matrix (CM) approach. The QLEs~\eqref{qleULB} can be rewritten in the quadrature form, where the quadrature fluctuations are defined as $\delta X_O = (\delta O + \delta O^{\dagger})/\!\sqrt{2}$ and $\delta Y_O = i (\delta O^{\dagger} - \delta O)/\!\sqrt{2}$ ($O = U, L, b_1, b_2$), which can further be cast in the following matrix form:
\begin{align}
	\begin{split}\label{dotu}
	\dot{u}(t)={\cal R}u(t) + n(t).
	\end{split}
\end{align}
Here, $u(t)=[\delta X_U(t),\delta Y_U(t),\delta X_L(t),\delta Y_L(t),\delta X_{b_1}(t),\delta Y_{b_1}(t),$ $\delta X_{b_2}(t),\delta Y_{b_2}(t)]^{\rm T}$ is the vector of quadrature fluctuations, $n(t)=[\sqrt[]{2\kappa_u}X_U^{in},\sqrt[]{2\kappa_U}Y_U^{in},\sqrt[]{2\kappa_l}X_L^{in},\sqrt[]{2\kappa_l}Y_L^{in}, \sqrt[]{2\kappa_1}X_{b_1}^{in},\sqrt[]{2\kappa_1}Y_{b_1}^{in},$ $\sqrt[]{2\kappa_2}X_{b_2}^{in},\sqrt[]{2\kappa_2}Y_{b_2}^{in}]^{\rm T}$ is the vector of input noises, written also in the quadrature form, and the drift matrix ${\cal R}$ is given by
\begin{widetext}
\begin{align}
	\cal R=\begin{pmatrix}
		-\kappa_u & {\Delta}_u & 0 & 0 & - G_{1M} \sin \theta &0 & - G_{2M} \sin \theta &0 \\
		-{\Delta}_u & -\kappa_u &0 & 0 & 0 & 0 & 0 & 0\\
		0 & 0 & -\kappa_l & {\Delta}_l & - G_{1M} \cos \theta &0 & - G_{2M} \cos \theta &0 \\
		0 & 0 &-{\Delta}_l & -\kappa_l & 0 & 0 & 0 & 0\\
		0 & 0 & 0 & 0 & -\kappa_1 & \omega_1 & 0 & 0\\
		0 & G_{1M} \sin \theta & 0 &  G_{1M} \cos \theta & -\omega_1 & -\kappa_1 & 0 & 0\\
		0 & 0 & 0 & 0 & 0 & 0 & -\kappa_2 & \omega_2\\
		0 & G_{2M} \sin \theta & 0 &  G_{2M} \cos \theta  & 0 & 0 & -\omega_2 & -\kappa_2\\
	\end{pmatrix}.
\end{align}
\end{widetext}

Since the input noises are Gaussian and the system dynamics is linearized, the steady state of the quadrature fluctuations is a four-mode Gaussian state, which can be fully characterized by an $8\times8$ CM $V$ with its entries defined as $V_{ij}=\frac{1}{2} \langle u_i(t)u_j({t}) + u_j({t})u_i(t) \rangle$ $(i,j=1,2,...,8)$. The steady-state CM can be achieved by directly solving the Lyapunov equation~\cite{Vitali07}:
\begin{align}
	\begin{split}
	{\cal R} V+V{\cal R}^{\rm T} = -D,
	\end{split}
\end{align}
where the diffusion matrix $D$ is defined via $D_{ij}\,\delta(t \,{-} \,t')=\langle n_i(t)n_j(t')+n_j(t')n_i(t) \rangle/2$, and takes the form
\begin{widetext}
\begin{align}
	D = \begin{pmatrix}
		\kappa_u(2n_u {+}\,1) & 0 & {\cal D} & 0 & 0 &0 & 0 & 0 \\
		0 & \kappa_u(2n_u {+}\,1) &0 & {\cal D} & 0 & 0 & 0 & 0\\
		{\cal D} & 0 & \kappa_l(2n_l {+}\,1) & 0 & 0 & 0 & 0 &0 \\
		0 & {\cal D} & 0 & \kappa_l(2n_l {+}\,1) & 0 & 0 & 0 & 0\\
		0 & 0 & 0 & 0 & \kappa_1(2n_{b_1}+1) & 0 & 0 & 0\\
		0 & 0 & 0 & 0 & 0 & \kappa_1(2n_{b_1}+1) & 0 & 0\\
		0 & 0 & 0 & 0 & 0 & 0 & \kappa_2(2n_{b_2}+1) & 0\\
		0 & 0 & 0 & 0 & 0 & 0 & 0 & \kappa_2(2n_{b_2}+1)\\
	\end{pmatrix}.
\end{align}
\end{widetext}
Here, the mean thermal excitation numbers of the two polaritons are defined as: $n_u =\frac{1}{2} \Big\{\big[\kappa_a \cos^2 \theta (2 n_a + 1) + \kappa_m \sin^2 \theta (2 n_m + 1) \big]/\kappa_u -1 \Big\}$ and $ n_l = \frac{1}{2} \Big\{ \big[\kappa_a \sin^2 \theta (2 n_a + 1) + \kappa_m \cos^2 \theta (2 n_m + 1)\big]/\kappa_l - 1 \Big\} $, and ${\cal D} = \frac{1}{2} \tan2\theta \, \big[-\kappa_u(2n_u +1)+\kappa_l(2n_l +1)\big]$.  

The steady-state mean energies of the two mechanical modes are related to the variances of the mechanical fluctuations, which can be extracted from the CM $V$, via 
\begin{align}
	\begin{split}
		U_1 &= \hbar \omega_1 \frac{\langle \delta X_{b_1}^2 \rangle + \langle \delta Y_{b_1}^2 \rangle}{2} \equiv \hbar \omega_1 \Big(n_{1,{\rm eff}} + \frac{1}{2} \Big),\\
		U_2 &= \hbar \omega_2 \frac{\langle \delta X_{b_2}^2 \rangle + \langle \delta Y_{b_2}^2 \rangle}{2} \equiv \hbar \omega_2 \Big(n_{2,{\rm eff}} + \frac{1}{2} \Big).\\
	\end{split}
\end{align}
Therefore, the effective mean phonon numbers are obtained
\begin{align}
	\begin{split}
		n_{1,{\rm eff}} &= \frac{\langle \delta X_{b_1}^2 \rangle + \langle \delta Y_{b_1}^2 \rangle -1}{2},\\ 
		n_{2,{\rm eff}} &= \frac{\langle \delta X_{b_2}^2 \rangle + \langle \delta Y_{b_2}^2 \rangle -1}{2}.
	\end{split}
\end{align}

\end{document}